# OPTIMIZATION OF RF PHASE AND BEAM LOADING DISTRIBUTION AMONG RF STATIONS IN SuperKEKB

S. Ogasawara[†], K. Akai, T. Kobayashi, K. Nakanishi, M. Nishiwaki,
High Energy Accelerator Research Organization (KEK), Tsukuba, Japan

*Abstract*

SuperKEKB is the e-/e+ collider which targets the world highest luminosity. In recent operation, SuperKEKB achieved a new world record of $4.71 \times 10^{34}\ cm^{-2}s^{-1}$ for luminosity with the beam current of 1.4 A. In the future, the beam current will be increased further to aim at the design value of 3.6 A and much higher luminosity. The RF system consists of 38 cavities (30 klystron stations), which share the huge beam loading brought by high current beam with each other cavities. For beam stability and power efficiency, it is important to distribute the beam loading properly among RF cavities. It is equivalent to adjust the acceleration phase of each cavity. However, it is difficult to evaluate the acceleration phase using only the pickup signal. Therefore, we established a method to evaluate the beam loading balance among RF stations from the RF power measurement for each cavity, and to adjust the acceleration phase. This report introduces the method for evaluating and optimizing the beam loading (acceleration phase) among stations in SuperKEKB, which has a large number of RF stations, and its operation.

## OVERVIEW

SuperKEKB is an e-/e+ collider upgraded from KEKB, which targets the world highest luminosity [1]. In recent operation, we achieved a new world record luminosity with the beam current of 1.4 A [2]. SuperKEKB consists of two rings, 7-GeV electron ring (High Energy Ring, HER) and 4-GeV positron ring (Low Energy Ring, LER). To achieve higher luminosity, the beam current is designed as a large value, 3.6 A (LER). Due to the huge beam current, the huge RF power (beam loading) is required.

The RF system consists of two-types 38 cavities in two rings [3,4]. The number of RF (klystron) stations is 30 with three types of station. In order to accumulate the high current beam stably and efficiently, it is important to optimize the beam loading balance among RF stations with considering the difference of station types. The optimization of the beam loading corresponds to adjust the station phase.

This paper introduces a method for evaluating and optimizing the beam loading among the RF stations in SuperKEKB.

## THE RF SYSTEM

Table 1 shows RF-related operation parameters. Figure 1 shows an arrangement of RF stations in SuperKEKB rings. There are three acceleration (RF) sections at each ring (total six sections for the two rings). The RF frequency is 508.9 MHz

---
† shunto.ogasawara@kek.jp

Table 1: Operation Parameters of RF System. [3-5]

|  | LER | HER |
|---|---|---|
| Beam Energy [GeV] | 4.0 | 7.0 |
| Beam Current [A] | 3.6* | 2.6* |
| Loss Energy [MeV/turn] | 1.76 | 2.43 |
| Cavity Type | ARES | SCC/ARES |
| RF Frequency [MHz] | 508.9 | |
| # of Cavities (# of Stations) | 22 (16) | 8(8)/8(6) |
| RF Peak Voltage [MV/cavity] | 0.5* | 1.5*/0.5* |
| Beam Power [kW/cavity] | 400~* | 400*/600* |
| Klystron Power [kW/station] | ~800* | ~450*/800* |

*: design value

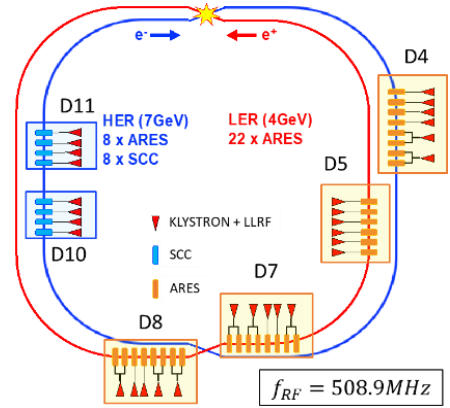

Figure 1: Arrangement of RF stations in SuperKEKB

SuperKEKB has two types of cavities: normal conducting "ARES" cavity and superconducting cavity (SCC) [6]. ARES cavity is a unique cavity system specialized for KEKB [7]. SCC is applied to only HER. HER has eight ARES cavities and eight SCCs. LER has 22 ARES cavities.

The RF system has three station types of "ARES-1:1", "ARES-1:2" and "SCC". ARES-1:1 means that one klystron (KLY) [8] drives one ARES cavity, and ARES-1:2 means that one KLY drives two cavities. SCC station has one cavity driven by one KLY. For the design of SuperKEKB, the 1:1 configuration will be applied to all ARES stations in HER. However, the present state is at the halfway stage of the upgrade for the design.

Figure 2 shows the RF system configuration of an acceleration section (example of the D07 section), including LLRF control systems [9]. "Section Phase" is the phase shifter for the RF reference for the section. And "Station Phase" is that for the cavity phase of each station. They are remote controllable. "Station Phase" is used to adjust the cavity phase (beam loading) for each station individually.

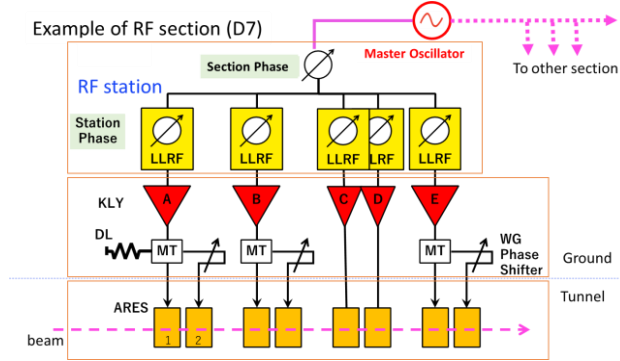

Figure 2: Phase shifters in RF D07 section.

# BEAM LOADING AND ACCERELATION PHASE

This section introduces basic theories related to beam loading adjustment. In SuperKEKB, the particle speed of the beam can be considered as the light speed.

## Beam loading and acceleration phase

For the case that the beam is accelerated by a cavity at phase $\phi_s$, the beam power $P_{beam}$ is given by

$$P_{beam} = V_c \cos \phi_s \cdot I_{beam}, \qquad (1)$$

where $V_c$ is the amplitude of the acceleration voltage of the cavity and $I_{beam}$ is the beam current. This is the power provided by the cavity to the beam. It is the beam loading for the cavity.

For ARES-1:2 station, the voltage and phase correspond to those of the vector sum voltage of the two cavities.

In a ring which has many RF stations like SuperKEKB, the vector sum voltage of all cavities (amplitude: $V_{c.sum}$, phase: $\phi_{s.sum}$) should be considered as the acceleration voltage for the ring. Figure 3 shows an example of the vector sum of three stations. The horizontal axis corresponds to the beam phase. In Fig. 3, the phase of "V3" is different from that of other stations. Dashed vector indicates the vector sum of case that the phase of all stations in the same.

$\phi_{s.sum}$ is determined by $V_0$ which corresponds to One-Turn loss $U_0$ ($V_0 \equiv U_0/e$) as follows.

$$V_0 = V_{c.sum} \cos \phi_{s.sum}. \qquad (2)$$

And then, $\phi_s$ of each station is consequently determined. If phases are not uniform among stations, the vector sum amplitude becomes smaller, and also, $\phi_{s.sum}$ and restoring force of synchrotron oscillation become smaller. Therefore, it is better to equalize RF phases among stations basically. The phase equalization is equivalent to uniform beam loadings among stations.

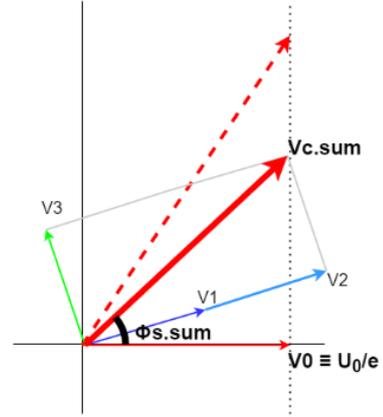

Figure 3: Example of vector sum of 3 stations.

## Loading distribution according to station type

All stations are operated by using the same model of KLY which the related power is 1 MW. However, we have three types of RF stations having different characteristics. For example, in ARES 1:2 station, KLY has to provide 2 times larger power than ARES-1:1 station. For SCC station, the practical acceptable power is limited to about 400 kW due to the achievement of the input coupler [6]. If the beam loading is uniformed among stations without considering of these station features, the design beam current cannot be achieved due to the RF power limit as described below.

Figures 4, 5 and 6 show estimations of required RF power for three station types of HER. For the KLY output power, the operation maximum is practically limited to about 800 kW due to the linearity property for stable $V_c$ regulation control.

Figure 4 shows a case of uniform beam loading for all stations. In this case, the SCC input power achieves the practical limit at the beam current of 1.5 A. Figure 5 shows a case of applying an appropriate phase difference among the station types. In this case, the reachable beam current is increased to 2.2 A, although it is slightly insufficient for the design current of 2.6 A. Figure 6 shows the case of the completed upgrade of the RF configuration (all ARES 1:2 stations become 1:1) as the design. Even this case, the beam loading optimization is still important.

Additionally, in the actual operation for phase adjustment, the hardware condition or other individual difference should be concerned. In some cases, specified station phase may be required to be fixed to arbitrary value. It would be useful to care malfunctioning (unhealthy) cavity.

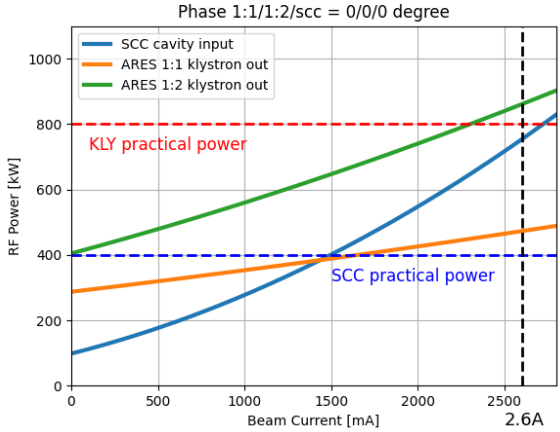

Figure 4: Estimation of required RF power with respect to beam current in HER, case of no phase difference.

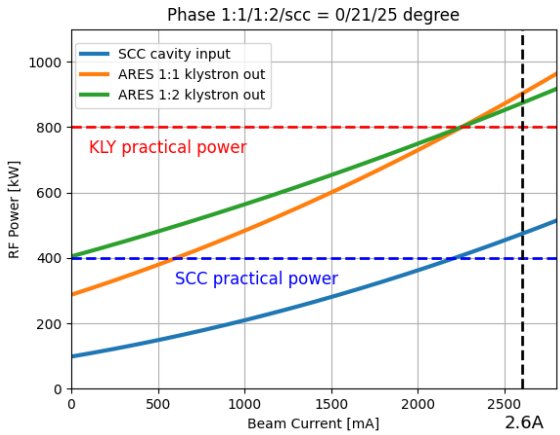

Figure 5: Case of applied appropriate phase difference.

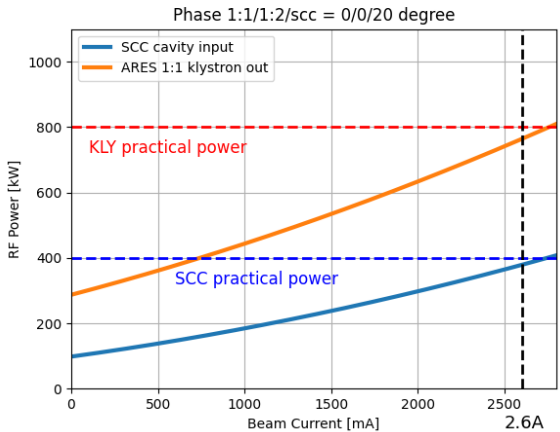

Figure 6: Case of 1:1 configuration for all ARES stations.

### Requirements for phase adjustment

One more important requirement should be noted for the adjustment. The beam phase must be kept through the adjustment. If the beam phase changes in either ring, the longitudinal bunch position will shift. Then, the collision point will be displaced from the optimum (ideal) position, and the luminosity will be lost. According to recent experience of beam tuning, when the beam phase changed by 1°, the luminosity will be decreased by about 5-8%. Although $V_{c.sum}$ and $\phi_{s.sum}$ are changed by the phase adjustment, the relationship of Eq. (2) must be kept to maintain the luminosity. Therefore, in the phase adjustment, the acceleration phase must be chosen so that satisfy Eq. (2).

Additionally, the adjustment work is preferred to be repeated periodically to compensate phase drift. Hence, it is also important that the adjustment method should be easy way.

In summary, the adjustment has following requirements.
- Arbitrary phase difference can be applied among stations.
- In arbitrary station, phase can be also specified to arbitrary value.
- Uniform phases among stations which have the same condition.
- Do not change beam phase by applying the adjustment.
- The adjustment method has to be easy way.

### Beam loading adjustment method

To change the beam loading, by controlling phase shifters shown in Fig. 2, the cavity phase is changed relative to the beam. At the startup of an accelerator, a cavity pickup signal is measured directly and compared to neighboring cavities. However, it is not easy to compare RF phase among RF sections located at 1-km distance from each other. And such measurement work is also impossible during beam operation.

One of the other possible methods is to observe the synchrotron oscillation frequency. In this method, we should scan phase shifters for all stations sequentially and should find the phase maximizing the synchrotron frequency. So, this method takes much time and effort. And it is not easy to apply accurate phase difference. Thus, this method doesn't satisfy the requirements.

Besides on the above consideration, we adopted a method that using RF power measurement. In this method, the ideal phase setting of all stations can be simultaneously determined at once. And this method does not disturb the beam operation.

As a side note, in this paper, it is assumed that the beam loading balance between the two cavities of ARES-1:2 station is equalized. The method of equalization of beam loading for the two cavities of ARES 1:2 station has been established [10].

# EVALUATION OF BEAM LOADING

## Evaluation of acceleration phase

As a first step, we should know an acceleration phase $\phi_s$ for each station. The RF powers interchanged through a cavity has following relationship:

$$P_{kly} - P_{ref} = P_{wall} + P_{beam}, \quad (3)$$

where $P_{kly}$ is the KLY output power (cavity feeding power), $P_{ref}$ is the cavity reflection power, and $P_{wall}$ is the cavity wall loss. Coupling factor of the input coupler is set so that $P_{ref}$ vanishes at the design beam current. $P_{wall}$ is proportional to $V_c^2$. Figure 7 shows relationship between these RF powers and the beam current.

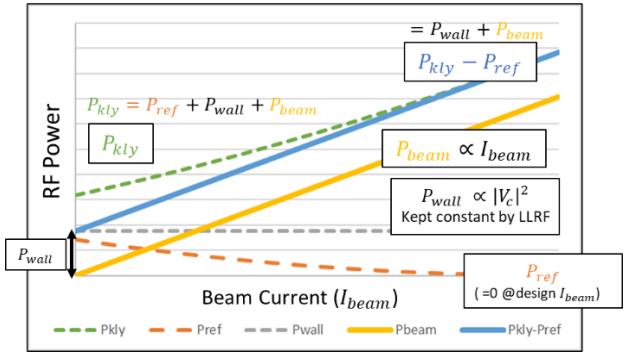

Figure 7: RF powers with respect to beam current.

$P_{kly}$ and $P_{ref}$ are directly measurable. $P_{wall}$ is calculatable from Q-value and cavity pickup, but it is also obtainable easily from $P_{kly} - P_{ref}$ where $I_{beam} = 0$.

$P_{beam}$ can be obtained from these RF powers and then $\phi_s$ is given from Eq. (1). However, high power RF measurement is not accurate enough. Therefore, we use linear fitting of the $P_{beam}$ plot shown in Fig. 7 measured during stacking beam current. $\phi_s$ is given by the slope of the linear fitting of $P_{kly} - P_{ref}$. The intercept of the plot of $P_{kly} - P_{ref}$ gives $P_{wall}$ as shown in Fig. 7.

## Calculation of target phase

In this subsection, followings are assumed:
- There are $N$ stations in a ring.
- $V_i$ and $\phi_i$ are the cavity voltage and the station phase, respectively, for the $i$-th station.
- The beam current and all $V_i$s are known.
- Each $\phi_i$ has been also obtained by the method explained in the previous subsection.

Under the conditions above, $V_{c.sum}$ and $\phi_{s.sum}$ can be calculated. And $V_0$ is also calculatable from Eq. (2).

First, if station phases are equalized into the target phase $\phi_{tgt}$, to keep the beam phase,

$$V_0 = \left(\sum_i^N V_i\right) \cos\phi_{tgt} \quad (4)$$

should be satisfied. For applying phase difference $\Delta\phi_i$ at $i$-th station, $\phi_{tgt}$ has to satisfy

$$V_0 = \sum_i^N V_i \cos(\phi_{tgt} + \Delta\phi_i). \quad (5)$$

Additionally, in case of specifying station phase of $j$-th station to $\phi_{j\_tgt}$ arbitrary, $\phi_{tgt}$ are given by

$$V_0 - V_j \cos\phi_{j\_tgt} = \sum_{i \neq j}^N V_i \cos(\phi_{tgt} + \Delta\phi_i). \quad (6)$$

For specification of arbitrary phases for multiple stations, summation for $j$ is considered for the second term in the left-hand side of Eq. (6).

# THE BEAM LOADING OPTIMIZATION TOOL

## Development of optimization tool

A software (optimization tool) is developed for phase evaluation and optimum phase calculation by using Python3. Figure 8 shows the screenshot of the tool.

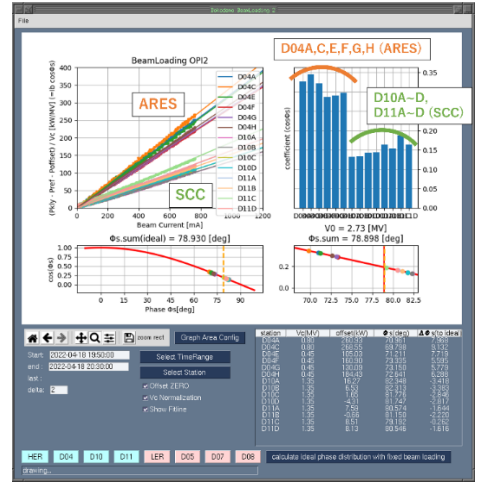

Figure 8: Example of beam loading optimization tool.

The tool has following functions:
- Show $P_{beam} - I_{beam}$ plot for specified period.
- Deduct offset ($P_{wall}$) from the plot.
- Normalize $P_{beam}$ by $V_c$ for comparison among stations having different $V_c$.
- Show slope of plot by bar graph for each station.
- Show phase distribution for each station.
- Calculate optimum phase (correction amount for phase shifter) according to Eq. (5).

Figure 8 shows the beam loading in HER on one day before optimization with offset deduction and $V_c$ normalization. The D04A and D04C stations are ARES 1:2 type. D04E, F, G and H are ARES 1:1 type. D10 and D11 stations are SCC stations. In this situation, the phase difference of about 10° between ARES and SCC stations is applied. However, the phase difference among ARES stations is not clear. And among SCC stations, the phases are not uniform.

*Example of phase adjustment*

For the situation of Fig. 8, the beam loading adjustment was performed. The optimum phase settings are calculated by the calculation tool as shown in Fig. 9. The calculation tool is opened by a button at the bottom right of Fig. 8. In this case, optimum phases are calculated by giving following specification:
- Apply the phase difference of 4° between ARES 1:1 and 1:2 stations.
- Set all SCC station phases to 81° as a fix value.

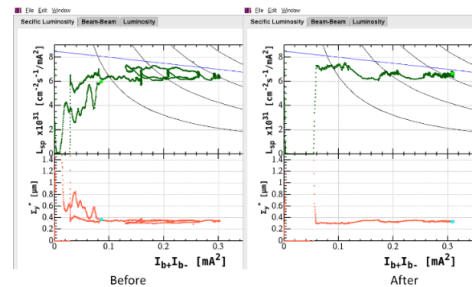

Figure 9: Optimum phase calculation tool.

Figure 10 shows the result of the adjustment. The calculation and adjustment were iterated twice to improve accuracy. After the adjustment, we can see three lines for three station types clearly. Most station phases were getting the target phase within error of 0.5°.

Figure 11 shows the specific luminosity (luminosity normalized by the bunch current) before and after the adjustment. No significant change was found in the luminosity.

## SUMMARY

The beam loading (acceleration phase) optimization method in SuperKEKB has been introduced. The evaluation can be conducted by a simple procedure using only RF power measurement and other parameters that are easy to know. This optimization is essential to store high current beam. The phase adjustment considering hardware condition can reduce the frequency of hardware trouble and beam abort, and contribute to stable operation of the accelerator. Additionally, easy obtaining of the synchronous phase values is useful for various beam studies.

For the next step, we want to automate this procedure to keep ideal beam loading distribution during long term operation of the beam collision for physics run with high luminosity.

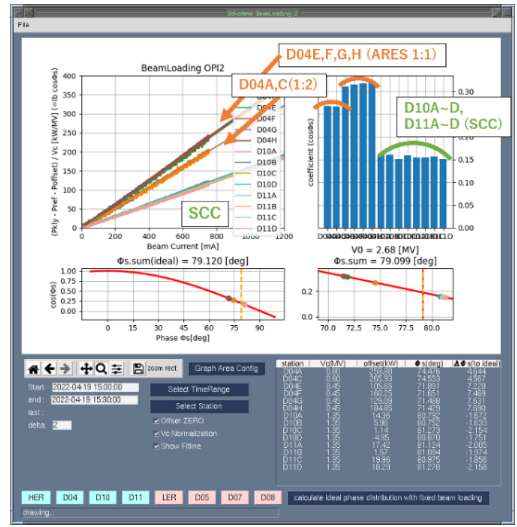

Figure 10: Result of the adjustment.

Figure 11: Specific luminosity before and after the adjustment.

## REFERENCES

[1] Y. Ohnishi *et al.*, "Accelerator design at SuperKEKB", Prog. Theor. Exp. Phys., vol. 2013, no. 3, pp.03A011, 2013

[2] Operation status and plan, https://www-linac.kek.jp/skekb/status/web/status_plan.md.html

[3] K. Akai et al., "RF System for SuperKEKB", in *Proc. PASJ2010*, Himeji, Japan, Aug. 2010, pp.177-181.

[4] M. Nishiwaki, "RF system (1)", OHO'19 (2019)

[5] T. Kobayashi, "RF system (2)", OHO'19 (2019)

[6] M. Nishiwaki et al., "Status of Superconducting Accelerating Cavity and Development of SiC Damper for SuperKEKB", in *Proc. PASJ2017*, Sapporo, Japan, Aug. 2017, pp. 914-918.

[7] T. Kageyama, et al., "Development of High-Power ARES Cavities", in *Proc. PAC97*, Vancouver, Canada, May. 1997, pp. 2902-2904.

[8] K. Watanabe et al., "Current Status of the High-Power RF Systems during Phase2 Operation In SuperKEKB", in *Proc. PASJ 2018*, Nagaoka, Japan, pp.464-467.

[9] T. Kobayashi et al., " Operation Status of LLRF Control System in SuperKEKB", in *Proc. PASJ2021*, Online, Aug. 2021, TUP044

[10] T. Kobayashi et al., "Phase adjustment between cavities with beam loadings disparity in high power RF distribution system at SuperKEKB", in *Proc. PASJ2020*, Online, Sep. 2020, WEPP38